\documentstyle[preprint,aps,prb]{revtex}
\title{A geometrical method towards first integrals for dynamical systems}
\author{Simon Labrunie and Robert Conte\\
        Service de physique de l'\'etat condens\'e\\
        CEA Saclay\\
        91191 Gif-sur-Yvette cedex (France)\\
        e-mail: {\tt labrun@spec.saclay.cea.fr}}

\newcommand{\s}{\sigma}
\renewcommand{\a}{\alpha}
\renewcommand{\b}{\beta}
\renewcommand{\r}{\varrho}
\renewcommand{\S}{\Sigma}

\newcommand{\cte}{{\mathop{\hbox{{\rm cst}}}\nolimits}}
\newcommand{\e}{{\mathop{\hbox{{\rm e}}}\nolimits}}
 
\let \\=\partial

\newcommand{\eqn}{equation}
\newcommand{\Eqn}{Equation}
\newcommand{\sys}{system}
\newcommand{\dsys}{dynamical \sys}
\newcommand{\fil}{first integral}
\newcommand{\pol}{polynomial}
\newcommand{\dpol}{Darboux \pol}
\newcommand{\deno}{denominator}
\newcommand{\dif}{differential}
\newcommand{\deqn}{\dif\ \eqn}
\newcommand{\sq}{following}
\newcommand{\wrt}{with respect to}
\newcommand{\st}{such that}

\begin{document}
\maketitle

\begin{abstract}
We develop a method, based on Darboux' and Liouville's works, to find first integrals and/or invariant manifolds for a physically relevant class of dynamical systems, without making any assumption on these elements' form. We apply it to three dynamical systems: Lotka--Volterra, Lorenz and Rikitake.
\end{abstract}

\section{Historical overview.\label{OVERVW}}

In~\cite{liu,tr2}, Roger Liouville and A.~Tresse developed a method for deciding whether two \deqn s of the form
\begin{equation}
{d^2y \over dx^2} + a_1(x,y)\,\left({dy \over dx}\right)^3 + 3\,a_2(x,y)\,\left({dy \over dx}\right)^2 + 3\,a_3(x,y)\,{dy \over dx} + a_4(x,y) = 0
\label{liu}
\end{equation}
where the $a_i$ are arbitrary functions of the real or complex variables $x$ and $y$, are geometrically equivalent, i.e.~can be transformed into each other by the most general dependent and independent variable change
\begin{equation}
x'=\varphi(x,y),\quad y'=\psi(x,y)
\label{tsf}
\end{equation}
This method was based on the construction of a ``relative invariant" function called $\nu_5$ of the $a_i$ and of their derivatives, \st\ in any transformation~(\ref{tsf}) it becomes $\nu'_5 = J(x,y)^{-5}\,\nu_5$ where $J(x,y)$ is the Jacobian of the transformation. In the general case, two \eqn s \st\ their $\nu_5$ are non-zero and proportional to each other are indeed equivalent. If $\nu_5=0$ for both, however, one cannot conclude at first, and other invariants, involving higher derivatives of the $a_i$, must be calculated in order to decide. As an application, Liouville proposed the effective reduction of \Eqn~(\ref{liu}) into its simplest canonical form, which in most cases leads to an explicit integration.

\medbreak

Here we will adopt another point of view. We have derived from these theories a method for finding out \fil s for a wide and physically important class of \dsys s without having to make any ansatz on their functional form. In the rest of this section, we shall recall some mathematical results of Darboux~\cite{dbx}, Liouville and Tresse.  Then we explain our method in Section~\ref{SHORT}, and apply it in Section~\ref{RESULTS} to three well-known \dsys s. Finally, Section~\ref{CONCL} discusses our results summarised in Table~\ref{table1}.

\subsection{Essentials of Liouville theory.}

Consider a \deqn\ like~(\ref{liu}). Liouville~\cite{liu} defined the \sq\ functions --- which are seen as functions of $(x,y)$, forgetting the supposed relation between those variables.
\begin{eqnarray}
L_2 &=& {\partial \over\partial x} \left( {\partial a_1\over\partial x} -3\,a_1\,a_3 \right) + {\partial \over\partial y} \left( {\partial a_3\over\partial y} - 2\,{\partial a_2\over\partial x} + a_1\,a_4 \right) \nonumber\\
     &-&             3\,a_2\,\left( {\partial a_3\over\partial y} - 2\,{\partial a_2\over\partial x} + a_1\,a_4 \right) + a_1\,\left( {\partial a_4\over\partial y} + 3\,a_2\,a_4 \right) \label{L2}\\
L_1 &=& {\partial \over\partial y}\left( {\partial a_4\over\partial y} + 3\,a_2\,a_4 \right) -{\partial \over\partial x}\left(2\,{\partial a_3\over\partial y} - {\partial a_2\over\partial x} +a_1\,a_4\right)\nonumber\\
    &-&              3\,a_3\,\left(2\,{\partial a_3\over\partial y} - {\partial a_2\over\partial x} +a_1\,a_4\right) -a_4\,\left({\partial a_1\over\partial x} -3\,a_1\,a_3 \right) \label{L1}\\
\nu_5 &=& L_2\,\left( L_1\,{\partial L_2\over\partial x} - L_2\,{\partial L_1\over\partial x}\right) + L_1\,\left(L_2\,{\partial L_1\over\partial y} - L_1\,{\partial L_2\over\partial y} \right)\nonumber\\
      & &-a_1\,L_1^3 + 3\,a_2\,L_1^2\,L_2 - 3\,a_3\,L_1\,L_2^2 + a_4\,L_2^3 \label{nu5}
\end{eqnarray}
The equation $\nu_5=0$ means that 
\begin{equation}
L_1\,dx+L_2\,dy=0
\label{subeq:gene}
\end{equation}
defines a particular solution of \Eqn~(\ref{liu}). We shall call~(\ref{subeq:gene}) a {\it sub\eqn\/} of \Eqn~(\ref{liu}), i.e.~(\ref{liu}) is a \dif\ consequence of~(\ref{subeq:gene}). Notice that in the case $L_2\equiv0$ the solution is {\it not\/} $L_1\equiv0$ --- which would mean an unexpected lowering of this \eqn's \dif\ order --- but $dx=0$, an absurdity as $x$ is seen as the independent variable. Similarly if $L_1\equiv0$, the solution is $dy=0$, a solution possibly present in \Eqn~(\ref{liu}) but not very interesting.

\smallbreak

Now suppose neither $L_1$ nor $L_2$ vanish identically and define $\a=-L_2/L_1$. Then
\begin{equation}
\nu_5 = L_1^3\,\left(\a\,{\\\a\over\\y} + {\\\a\over\\x} + a_1\,\a^3 + 3\,a_2\,\a^2 + 3\,a_3\,\a + a_4 \right)
\label{nu5:alpha}
\end{equation}
and \Eqn~(\ref{subeq:gene}) can be rewritten $dy/dx = \a(x,y)$. Conversely suppose there is a first-order sub\eqn\ to \Eqn~(\ref{liu}), namely $dy/dx = A(x,y)$.  Then $A$ is a solution to the first-order non-linear PDE
\begin{equation}
A\,{\\A\over\\y} + {\\A\over\\x} + a_1\,A^3 + 3\,a_2\,A^2 + 3\,a_3\,A + a_4 =0
\label{pde:A}
\end{equation}
we shall discuss later. Liouville theory is a finite effort tool for finding particular solutions to \Eqn~(\ref{pde:A}).

\subsection{\dpol s and \fil s for \pol\ \dsys s.}

Consider an autonomous \pol\ \dsys 
\begin{equation}
\dot x_i = V{}_i(x),\quad i=1\dots n
\label{pol:gene}
\end{equation}
We say that a \pol\ $f(x_1,\dots,x_n)$ is a {\it \dpol\/}~\cite{dbx} of~(\ref{pol:gene}) if there exists  a \pol\ ``eigenvalue" $p$ \st
$${df\over dt}\equiv\sum_{i=1}^n V{}_i\,{\\f\over\\x_i} = p\,f$$
In other words, there is an algebraic variety, defined by $f(x_1,\dots,x_n) = 0$, which is invariant by the flow of~$V$. In this respect, this notion is a neighbour of the notion of sub\eqn\ we have seen.

\medbreak

\dpol s are tools for finding out~\cite{man1,man2}, but also proving the non-existence (cf.~\cite{lab} for an example) of \fil s to \pol\ \dsys s. We shall not enter into the details. Let us just notice that a \pol\ \fil\ is simply a \dpol\ with eigenvalue~$0$; and that a \dpol\ $f$ with constant eigenvalue~$\a$ gives rise to the time-dependent \fil\ $f\,\e^{-\a\,t}$. More rational and algebraic \fil s can be built with the ``basic blocks" of \dpol s~\cite{man1,man2}; and, conversely, a theorem of Bruns~\cite{kum} says there cannot be an algebraic \fil\ of~(\ref{pol:gene}) unless there is a rational one, which in turn implies the existence of \dpol s.

\smallbreak

In brief, the problems of existence of \fil s and \dpol s for a \pol\ \dsys\ are very tightly linked. Notice also that all these objects, like ordinary eigenvectors and -values of linear endomorphisms, naturally live in $\hbox{{\bf C}}$.

\section{Principle of the method.\label{SHORT}}

We shall draw our interest to autonomous three-dimensional \pol\ \dsys s which are of {\it first degree\/} in one of their three variables, e.g.~$z$. Their general form is thus:
\begin{equation}
\left\{\matrix{\dot x = & V{}_x^0(x,y) + z\,V{}_x^1(x,y) \hfill \cr
         \dot y = & V{}_y^0(x,y) + z\,V{}_y^1(x,y) \hfill \cr
         \dot z = & V{}_z^0(x,y) + z\,V{}_z^1(x,y) \hfill \cr}
\right.
\label{deg1z:gene}
\end{equation}
which we may abbreviate as $\dot X = V(X)$. Dynamical \sys s of this kind are frequently met in physics: well-known examples are the Lorenz model, or the various three-wave interaction problems (Rabinovich etc.). Very often they are indeed of first degree in {\it all\/} their variables. We can use this feature for harvesting more information --- an example is given in the paragraph about the Lorenz model.

\medbreak

We assume to have found out and studied all solutions with $x=\cte$. Assuming $x$ nonconstant, we shall transform the \sys~(\ref{deg1z:gene}) into a non-autonomous second-order \deqn\ linking $y$ and $x$ which will turn out to be of type~(\ref{liu}). 

\smallbreak

Now we settle in a region of space where $\dot x\ne0$ and take $x$ as the independent variable, parametrising the integral curves of~(\ref{deg1z:gene}). The relation
\begin{equation}
\left(V{}_x^0(x,y) + z\,V{}_x^1(x,y)\right)\,{dy \over dx} = V{}_y^0(x,y) + z\,V{}_y^1(x,y)
\label{sing1}
\end{equation}
is satisfied along all integral curves. Hence, writing $p=dy/dx$,
\begin{equation}
z\,\left( V{}^1_x\,p - V{}^1_y\right) = V{}^0_y - p\,V{}^0_x
\label{sing2}
\end{equation}
These \eqn s define the mappings $\phi\colon(x,y,z)\mapsto(x,y,p)$ and $\phi^*\colon(x,y,p)\mapsto(x,y,z)$, which are homographic and hence:
\begin{enumerate}
\item They are one-to-one wherever they are defined and their determinant $C(x,y)=V{}^0_x\,V{}^1_y - V{}^0_y\,V{}^1_x$ is non-zero; as it involves only the variables $x$ and $y$ the surface $\S=\{C(x,y)=0\}$ can be seen either as a submanifold in the $(x,y,z)$ space or in the $(x,y,p)$ space.
\item The surfaces $S_1 = \left\{ V{}_x^0(x,y) + z\,V{}_x^1(x,y)=0 \right\}$ in the $(x,y,z)$ space and $S_2 = \left\{ V{}^1_x(x,y)\,p - V{}^1_y(x,y)=0 \right\}$ in the $(x,y,p)$ space are singular. Any point on $S_1\setminus\S$ is sent to $p=\infty$: this happens when $\dot x=0$, and the tangent to the integral curve is orthogonal to the $x$-axis, i.e.~``vertical" in $(x,y)$ representation ($dy/dx=\infty$). Similarly any point on $S_2\setminus\S$ is sent to $z=\infty$.
\item On $\S$, $\phi$ and $\phi^*$ are ``constant along fibres", i.e.~two points of $\S$ having different $z$ (or~$p$) are sent to the same image, having the same $(x,y)$ as the original point, hence lying on $\S$. Thus it is always possible to get it as the image of a point on $\S\setminus S_1$ and calculating it that way shows that $\phi(\S)=\S\cap S_2$; and similarly $\phi^*(\S)=\S\cap S_1$.
\end{enumerate}
Since we are concerned with a \dif\ problem, we have to study what the vector field $V(X)$ becomes under the action of the tangent map $T_X\phi$. And, indeed, points $X$ on $\S$ differing only in the $z$ coordinate have the same image by $\phi$, but different $V(X)$ \st\ the corresponding $T_X\phi(V(X))$ also generally differ. As all these vectors are attached to the common image of the points $X$, this can cause a loss of information, leading, as we shall see, to important practical difficulties.

\medbreak
 
Similarly, we find that
$$\left(V{}_x^0(x,y) + z\,V{}_x^1(x,y)\right)\,{dz \over dx} = V{}_z^0(x,y) + z\,V{}_z^1(x,y)
$$
We calculate $dz/dx$ by differentiating~(\ref{sing2}) \wrt\ $x$, putting the result into the previous formula, and then replacing $z$ itself with its value in function of $p$ given by \Eqn~(\ref{sing2}). This leads to a \deqn\ in $p$ which, in Cauchy form, reads
\begin{equation}
{dp\over dx} = {N(x,y,p)\over C(x,y)^2}
\label{gerbe}
\end{equation}
where $N$ is \pol\ in $(x,y,p)$ and of degree three in $p$. Interpreting $p$ as $dy/dx$, we see \Eqn~(\ref{gerbe}) as a \deqn\ of Liouville type like~(\ref{liu}).
There are two essential facts in these computations. One is that the \deno\ of~(\ref{gerbe}) is exactly the square of the determinant $C(x,y)$ of $\phi$, so \Eqn~(\ref{gerbe}) will not set any further problem as long as its construction is valid. The other one is that \Eqn s~(\ref{deg1z:gene}) are of degree one in $z$: it ensures not only the good behaviour of the $z\leftrightarrow p$ correspondence but also the Liouville form of the \deqn~(\ref{gerbe}).

\medbreak

We intend to apply Liouville theory to \Eqn~(\ref{gerbe}) in order to obtain sub\eqn s for it. Now, we must take care of their possible relationships with the forbidden surfaces. If a sub\eqn\ defines a curve in the $(x,y,p)$ space which is not contained in $\S$ or $S_2$, there is no problem: it will be pulled back into the $(x,y,z)$ space by the $\phi^*$ map, which coincides then with the reciprocal of $\phi$.

\smallbreak

But in the computation of \Eqn~(\ref{gerbe}), we have used the $z=\phi^*(x,y,p)$ map, and then suppressed the denominator $V{}^1_x\,p - V{}^1_y$. Thus, the singular manifold at $S_2$ has disappeared in~(\ref{gerbe}). But consider a curve plotted on $S_2$ (i.e.~$V{}^1_x\,p - V{}^1_y\equiv0$) which is, moreover, a {\it jet\/} (i.e.~$p\equiv dy/dx$). Then, identically
$${dy\over dx} = A(x,y) = {V{}^1_y(x,y)\over V{}^1_x(x,y)}$$
We can check that this $A$ is always a solution to \Eqn~(\ref{pde:A}), whatever the vector field $V$ may be. Hence {\em any jet plotted on $S_2$ is a sub\eqn\ of \Eqn~(\ref{gerbe})\/}. However, this jet cannot yield an invariant manifold in the $(x,y,z)$ space unless it is made of images by $\phi$ of points in this space. Now, it is easy to show that the only points on $S_2$ that can be written as $\phi(x,y,z)$ are those on $\S\cap S_2$; therefore, as we have seen, they are images of points also lying on $\S$.

\smallbreak

Yet we know that the pullback of the vector field $V$ is not necessarily well-behaved on $\S$. Thus, the \sys~(\ref{deg1z:gene}) and the \deqn~(\ref{gerbe}) can behave quite differently on $\S$.

If we find as \Eqn~(\ref{subeq:gene}) the \eqn\ of a jet on $S_2$, we have to check independently whether $\S$ is an invariant manifold for $V$ or not. In \dsys s containing parameters, this can be rephrased as: {\it find at what condition on the \sys's parameters the \eqn\ \deno\ $C(x,y)$ is a \dpol\/ for the \sys~(\ref{deg1z:gene})}.

\section{Results for several dynamical systems.\label{RESULTS}}

We have applied the method exposed in Section~\ref{SHORT} to three different \dsys s of type~(\ref{deg1z:gene}) depending on real parameters: Lotka--Volterra, Lorenz, Rikitake. We shall discuss the results obtained in the rest of this section.

\subsection{The $(a,b,c)$ Lotka--Volterra \sys.}

This remarkably symmetric \sys
\begin{equation}
\left\{\matrix{\dot x &=& x\,(c\,y+z) \hfill \cr
         \dot y &=& y\,(a\,z+x) \hfill \cr
         \dot z &=& z\,(b\,x+y) \hfill \cr}
\right.
\label{lotka}
\end{equation}
appeared first as a model for three-species competition, yet has been found later in plasma physics. A considerable amount of research has been done on it, using many techniques~\cite{lab,bou,str,gra}. Here we shall follow the process exposed in Section~\ref{SHORT}.

\smallbreak

Since \Eqn s~(\ref{lotka}) are invariant by simultaneous circular permutations of $(x,y,z)$ and $(a,b,c)$, it is equivalent to perform the method with any couple of variables. Once this is done, more results can be got by the above symmetry. There is also a symmetry in taking $x=x'/b,\ y=z'/c,\ z=y'/a$ and $a=1/c',\ b=1/b',\ c=1/a'$, which will appear in the distribution of the $\nu_5=0$ cases.

So, we take $z$ as the independent variable and eliminate $x$, and find
\begin{equation}
L_1 = {(b-1)(1 + a\,b\,c)\,Q^1_{abc}(y,z)\over z\,(y-a\,b\,z)^4}
\label{L1:lotka}
\end{equation}
and
\begin{equation}
L_2 = {(b-1)(1 + a\,b\,c)\,Q^2_{abc}(y,z)\over z\,(y-a\,b\,z)^4}
\label{L2:lotka}
\end{equation}
and
\begin{equation}
\nu_5 = {{(b-1)^3\,(1 + a\,b\,c)^3\,P_{abc}(y,z)}\over {z^2\,y^2\,(y-a\,b\,z)^{10}}}
\label{nu5:lotka}
\end{equation}
where $P_{abc},\ Q^1_{abc},\ Q^2_{abc}$ are \pol s whose coefficients depend on $(a,b,c)$ and which we do not write down for the sake of brevity. The cases $1 + a\,b\,c=0$ and $b=1$ are known: the first one is the full integrability of the \sys, with, in particular, the \fil\ $a\,b\,x + y - a\,z$; in the second one we have the \dpol\ $y-a\,z$ whose eigenvalue is $x$. We remark this \dpol\ is exactly the \deno\dots\ Notice also that in those two cases, $L_1$ and $L_2$ vanish together with $\nu_5$ so that \Eqn~(\ref{subeq:gene}) is an identity and cannot be used for finding out \dpol s.

\smallbreak

Now the cases where all coefficients of $P_{abc}$ are zero are listed below in Table~\ref{table1}.
We notice the presence of the symmetry $a=1/c',\ b=1/b',\ c=1/a'$ in this list; one case ($a = 1,\ b = 1,\ c = 1$) is self-symmetric. We shall handle in some detail one of these ``exotic" cases, viz.~$(a = 1/4,\ b = 2,\ c = -5)$. The sub\eqn~(\ref{subeq:gene}) reads:
\begin{equation}
y\left(-3\,z^2 + 16\,z\,y - 32\,y^2\right) + 4\,z\,{dy\over dz}\,\left( z^2 - 7\,z\,y + 16\,y^2 \right) = 0
\label{subeq:exo}
\end{equation}
Reverting to the original variables $(x,y,z)$ changes this \eqn\ in
\begin{equation}
y\,(z-2\,y)\,(16\,y^2 - 2\,x\,z - 8\,y\,z + z^2)=0
\label{inv:exo}
\end{equation}
The expression $z-2\,y$ is proportional to the equation \deno; its presence here is an artefact due, as we have seen, to a former suppression of \deno. It should not be taken in account since it corresponds to $b=1$. On the other hand, the other two factors are \dpol s, since their derivatives \wrt\ the \sys~(\ref{lotka}) are $\dot y = y\,(z/4+x)$ and
$${d\over dt}\,\left(16\,y^2 - 2\,x\,z - 8\,y\,z + z^2\right) = 2\,x\,\left(16\,y^2 - 2\,x\,z - 8\,y\,z + z^2\right)
$$
This illustrates the validity of the method in the general case. The results for all cases are summarised in Table~\ref{table1}. We get no information for the self-symmetric case since it is a specialisation of $b=1$, so our method cannot be applied.

\subsection{The Lorenz model.}

Another well-known and intensively studied~\cite{lor,seg,kus,lev,sen} \dsys, the Lorenz model
\begin{equation}
\left\{\matrix{\dot x& = & \s\,(y - x) \hfill \cr
         \dot y& = & \r\,x - y - x\,z \hfill \cr
         \dot z& = & x\,y - b\,z \hfill \cr}
\right.
\label{lorenz}
\end{equation}
originally thought as a simple model for atmospheric turbulence, was the first example of a low-dimensional chaotic deterministic \dsys~\cite{lor}. All known \fil s have been obtained or reobtained by Ku\'s~\cite{kus}, using the non-decisive procedure of Carleman embedding. Here we shall recover some of them {\it methodically\/} by the means of Liouville theory.

\medbreak

Since there is no symmetry among variables here, we can proceed three times to the calculation of the $L_{1,2}$ and $\nu_5$ functions, eliminating each time one of the three variables. Indeed, the Liouville \eqn\ like~(\ref{gerbe}) contains no more information than the \dsys\ like~(\ref{deg1z:gene}) does, but it has more singularities; yet, as we have seen, these singularities often contain useful matter about the \dsys's invariant manifold structure.

\medbreak

Eliminating $z$ and choosing $x$ as the independent variable yields $C(x,y)=\s\,x\,(y-x)$. Neither $x=0$ nor $y-x=0$ can be interesting invariant manifolds, since both imply $\dot x=0$, so $x=\cte$ and $y=\cte$. This, in turn, also implies that $z$ is constant, and the manifold reduces to a fixed point.

\medbreak

The functions $L_2$ and $\nu_5$ are (cf.~\cite{dry})
\begin{equation}
L_2 = {1 + b + \s\over\s\,(y-x)^3}
\label{L2:lorenz:xy}
\end{equation}
and
\begin{equation}
\nu_5 = {(1+b+\s)\,P_{b\s\r}(x,y)\over \s^5\,x^2\,(y-x)^{10}}
\label{nu5:lorenz:xy}
\end{equation}
In the obvious case $1+b+\s=0$, we also have $L_2=0$. Thus, as explained in Section~\ref{OVERVW}, our choice of variables was a bad one. Looking for other cases, we only get three points in the $(b,\s,\r)$ space, viz.~$\left(b=0,\,\s=-1\right)$, $\left(b=2/3,\,\s=1/3\right)$ and~$\left(b=-16/5,\,\s=-1/5,\,\r=-7/5\right)$. They are specialisations either of known cases~\cite{kus} or of the $1+b+\s=0$ case. In those cases, we get sub\eqn s that do not give rise to \dpol s, i.e.~\eqn s that represent jets plotted on the surface $S_2$.

\medbreak

Now, if we eliminate $y$ and choose $x$ as the independent variable, we get as \eqn\ \deno\ $C(x,z)=b\,z - x^2$. Its derivative \wrt\ the \sys~(\ref{lorenz}) is
\begin{eqnarray}
{d\over dt}\left( b\,z - x^2 \right) &=& (b-2\,\s)\,x\,y - (b^2\,z - 2\,\s\,x^2)
\label{dC:lorenz:xz}\\
  &=& -2\,\s\,(b\,z - x^2) + (b-2\,\s)\,(x\,y - b\,z)\nonumber
\end{eqnarray}
The remainder is of degree one in $x$. Thus $C$ is a \dpol\ iff $b=2\,\s$; then the eigenvalue is $-2\,\s$, so $I=(x^2 - 2\,\s\,z)\,\e^{2\,\s\,t}$ is a \fil~\cite{seg}.

\medbreak

We have calculated $\nu_5$ and found
\begin{equation}
\nu_5 = {(b-2\,\s)(1+b+\s)P_{b\s\r}\over \s^5\,(b\,z - x^2)^{10}}
\label{nu5:lorenz:xz}
\end{equation}
$P_{b\s\r}$ being \st\ its coefficients never simultaneously vanish. In the case $b=2\,\s$, \Eqn~(\ref{subeq:gene}) yields $x - \s\,dz/dx =0$. This is the \eqn\ of a jet on $S_2$, and since we are in the good case, we find $\S$ as the invariant surface.

\smallbreak

When $1+b+\s=0$, \Eqn~(\ref{subeq:gene}) still yields $x - \s\,dz/dx =0$. But in this case, $\S$ is not invariant and we do not have a \dpol.

\medbreak

Finally, we have taken $z$ as the independent variable and eliminated $x$. We find $C(y,z) = b\,\r\,z - b\,z^2 - y^2$ and
\begin{eqnarray}
{dC\over dt} &=& x\,y\,\left( (b-2)\,\r + 2\,(1-b)\,z \right) + 2\,y^2 + 2\,b^2\,z^2 - b^2\,\r\,z
\label{dC:lorenz:zy}\\
 &=& -2\,(b\,\r\,z - b\,z^2 - y^2) + R(x,y,z)\nonumber
\end{eqnarray}
the remainder $R$ being of first degree in $y$; hence $C$ is a \dpol\ iff $b=1$ and $\r=0$. Then, $dC/dt = -2\,C$ and we get that way the \fil\ $I = (z^2 + y^2)\,\e^{2\,t}$~\cite{seg}.

\smallbreak

As for $\nu_5$, it is equal to
\begin{equation}
\nu_5 = {(1+b+\s)\,y\,P_{b\s\r}(y,z)\over (b\,\r\,z - b\,z^2 - y^2)^{10}}
\label{nu5:lorenz:zy}
\end{equation}
where $P_{b\s\r}(y,z)\equiv 0$ iff $b=1$ and $\r=0$. Let us handle first the latter case. In that case, $L_1 = -\s\,z\,y/(z^2 + y^2)^2$ and $L_2 = -\s\,y^2/(z^2 + y^2)^2$, so \Eqn~(\ref{subeq:gene}) simplifies as $z\,dz+y\,dy=0$. This is the \eqn\ of a jet on $S_2$, and we get $\S$ as invariant manifold.
 
\smallbreak

Now, in the case $1+b+\s=0$, there is another simplification in \Eqn~(\ref{subeq:gene}), namely $y\,L_1\equiv(z-\r)\,L_2$ and hence $(z-\r)\,dz+y\,dy=0$. But this is the jet on $S_2$, so we get no information in this case.

\subsection{The Rikitake dynamo.}

This \dsys~\cite{bou}
\begin{equation}
\left\{\matrix{\dot x &=&-\mu\,x &+& y\,(z+\b) \hfill \cr
         \dot y &=& -\mu\,y &+& x\,(z-\b) \hfill \cr
         \dot z &=& \a &-& x\,y \hfill \cr}
\right.
\label{riki}
\end{equation}
models the variation of the earth's magnetic field with time.

\medbreak

Let us take $x$ as the privileged variable and eliminate $z$. The \deno\ in \Eqn~(\ref{gerbe}) is $C(x,y) = \mu\,(y^2 - x^2) + 2\,\b\,x\,y$, and
\begin{equation}
\nu_5 = {\b^2\,\mu^2\,P_{\a\b\mu}(x,y)\over (\mu\,(y^2 - x^2) + 2\,\b\,x\,y)^{10}}
\label{nu5:riki:xy}
\end{equation}
where the coefficients of $P_{\a\b\mu}$ cannot vanish together unless $\b=0$ or $\mu=0$.

\medbreak

The derivative of $C(x,y)$ \wrt\ the \sys~(\ref{riki}) is
\begin{equation}
{dC\over dt} = (y^2-x^2)\,(2\,\b^2 - 2\,\mu^2) - 2\,\b\,x\,y + 2\,\b\,z\,(x + y)
\label{dC:riki}
\end{equation}
Assume $C$ is a \dpol\ of eigenvalue $P(x,y,z) = A(x,y) + z\,B(x,y)$. Then the identification of the $z$ terms in \Eqn~(\ref{dC:riki}) gives $B\,C = 2\,\b\,(x+y)$. Since $C$ is of second degree, this is impossible unless $\b=0$.

\smallbreak

When $\b=0$, \Eqn~(\ref{dC:riki}) reads $dC/dt=-2\,\mu\,C$, so $C$ is a \dpol\ of this \sys, which gives the \fil\ $(y^2-x^2)\,\e^{2\,\mu\,t}$. On the other hand,
$$L_1 = {4\,x^2\,y\over\mu^2\,(x^2-y^2)^2},\quad L_2 = {-4\,x\,y^2\over\mu^2\,(x^2-y^2)^2}$$
hence \Eqn~(\ref{subeq:gene}) becomes $-x\,dx+y\,dy=0$. This is a jet on $S_2$, but we are in the ``good" case, and we recover the \dpol\ $y^2-x^2$.

\smallbreak

If now $\mu=0$ then 
$$L_1 = {-\a\,(3\,x^4 + 2\,x^2\,y^2 + 3\,y^4)\over 4\,\b^2\,x^3\,y^4},\quad
L_2 = {\a\,(3\,x^4 + 2\,x^2\,y^2 + 3\,y^4)\over 4\,\b^2\,x^4\,y^3}$$
so the sub\eqn\ is once more $-x\,dx+y\,dy=0$, the jet on $S_2$. Hence we do not obtain any \dpol\ unless $\b=0$.

\medbreak

We have also performed the computations with the other two couples of variables. They have not given more information than the previous ones. 

\section{Conclusion.\label{CONCL}}

We have obtained, {\it by a methodic procedure\/}, numerous cases of \dpol s for the Lotka--Volterra \sys. This ``\dpol\ searcher" can be seen as an input to algorithms which need \dpol s, such as the Prelle--Singer procedure. Up to now, that procedure began with a systematic search, which obliged to set an {\it a priori\/} limit on the \pol's degree in all its variables~\cite{man1}. Some new results~\cite{man2} allow to refine the search by restricting the choice of the possible highest-degree homogeneous components of the tentative \dpol s, while speeding it up when the \sys's coefficients are rational numbers. Yet they are valid for \dsys s of dimension~$2$, and until now have no counterpart in dimension~$3$.

Our method has also reobtained some known \fil s for the Lorenz and Rikitake \sys s, though all cases have not been found, and despite the ``divergence enigma" we now explain.

\smallbreak

In both Lorenz and Rikitake \sys s, the divergence is a constant, respectively $-1-b-\s$ and $-2\,\mu$. In both cases, its vanishing triggers the vanishing of $\nu_5$, but also the reduction of \Eqn~(\ref{subeq:gene}) to a singularity from which no information can be extracted. However, the Rikitake dynamo possesses when $\mu=0$ a time-dependent \fil\ $I=x^2 - y^2 + 4\,\b\,z - 4\,\a\,\b\,t$; this kind of \fil\ cannot be detected by our method, since it does not arise from a \dpol, but from a \pol\ $f(x,y,z)$ \st\ $df/dt=\cte$. Such a \fil\ may exist only for \dsys s having a constant term, so there is no chance to find any for the Lorenz model. But there may be a \fil\ of some special kind --- indeed, numerical experiments exhibit a regular behaviour when $1+b+\s=0$.

\medbreak

Table~\ref{table1} summarises the results obtained by our method. Abbreviations are: DS for ``\dsys", FI for ``\fil", and DE for ``Darboux element", i.e.~a couple of \pol s $(f,p)$ \st\ $df/dt = p\,f$.

\clearpage

\begin{table}[p]
\footnotesize
\begin{tabular}{|l|c|c|c|l|}
\hline
DS & $\nu_5$ & Denominator & Parameters & \multicolumn{1}{c|}{Information obtained} \\ 
\hline
      &  &  & $(0,2,c)$ & DE: $(y;x)$ \\
      &  &  & $({1/(2\,b)},b,1)$ & DE: $(b\,x - z; y)$ \\
      &  &  & $(1,b,{2/b})$ & DE: $(x - c\,y; z)$ \\
      &  &  & $({1/4},2,-5)$ & DE: $(16\,y^2 - 2\,x\,z - 8\,y\,z + z^2; 2\,x)$ \\
Lotka & $(y,z)$:(\ref{nu5:lotka}) & $y-a\,b\,z$ & $(-{1/5},{1/2},4)$ & DE: $(100\,y^2 - 25\,x\,y + 40\,y\,z + 4\,z^2;x)$ \\
$(a,b,c)$ &  &  & $(1,2,-2)$ & DE: $(y^2 - x\,z - y\,z; 2\,x + z)$ \\
      &  &  & $(-{1/2},{1/2},1)$ & DE: $(-2\,x\,y + 2\,y\,z + z^2; x+y)$ \\
      &  &  & $(1,1,1)$ & none \\
      &  &  & $(-{1/2},0,1)$ & none \\
      &  &  & $(0,{2/3},1)$ & DE: $(3\,z - 2\,x; y)$ \\
\cline{2-5}
      & other & \multicolumn{3}{c|}{cyclic permutation of the above results} \\
\hline
       &  &  & $(-1-\s,\s,\r)$ & none \\
       & $(x,y)$:(\ref{nu5:lorenz:xy}) & $x-y$ & $({2/3},{1/3},\r)$ & none \\
       &  &  & $(-16/5,-1/5,-7/5)$ & none \\
\cline{2-5}
Lorenz & $(x,z)$:(\ref{nu5:lorenz:xz}) & $b\,z - x^2$ & $(2\,\s,\s,\r)$ & FI: $(x^2 - 2\,\s\,z)\,\e^{2\,\s\,t}$ \\
$(b,\s,\r)$ &  &  & $(-1-\s,\s,\r)$ & none \\
\cline{2-5}
       & $(z,y)$:(\ref{nu5:lorenz:zy}) & $b\,z\,(\r - z) - y^2$ & $(1,\s,0)$ & FI: $(z^2+y^2)\,\e^{2\,t}$ \\
       &  &  & $(-1-\s,\s,\r)$ & none \\
\hline
Rikitake & $(x,y)$:(\ref{nu5:riki:xy}) & $\mu\,(y^2 - x^2)$ & $(\a,0,\mu)$ & FI: $(y^2 - x^2)\,\e^{2\,\mu\,t}$ \\
 $(\a,\b,\mu)$ &  & \hfill $+ 2\,\b\,x\,y$ &  $(\a,\b,0)$ & none \\
\cline{2-5}
         & other & \dots &  \multicolumn{2}{c|}{nothing more} \\
\hline
\end{tabular}
\caption{Results obtained by our method}
\label{table1}
\end{table}

\end{document}